\documentclass[twocolumn,preprintnumbers, sort&compress, floatfix]{revtex4}
\usepackage{amsmath}
\usepackage{amsfonts}
\usepackage{amssymb}
\usepackage{graphicx}
\usepackage{bm}
\usepackage{hyperref}
\usepackage{color}
\def\be{\begin{equation}}
\def\ee{\end{equation}}
\def\bi{\bibitem}

\begin{document}
\title{Studying the Intervention of an Unusual Term in $f(T)$ Gravity via Noether Symmetry}
\author{‪Behzad Tajahmad$ $}
\email{behzadtajahmad@yahoo.com}
\affiliation{Faculty of Physics, University of Tabriz, Tabriz, Iran}

\begin{abstract}
As usual, we observe an unknown coupling function, i.e. $F(\varphi)$, with a function of torsion and also curvature, i.e. $f(T)$ and $f(R)$, generally depending on a scalar field. In $f(R)$ case, it comes from quantum correlations and other reasons. Now, what if beside this term in $f(T)$-gravity context, we enhance the action through another term which depends on both scalar field and its derivatives? In this paper, we have added such an unprecedented term in the generic common action of $f(T)$-gravity such that in this new term, an unknown function of torsion has coupled with an unknown function of both scalar field and its derivatives. We have explained why we can append such term, in the introduction in details. By the use of Noether symmetry approach, we have considered its behavior and effect. We have shown that it does not produce an anomaly, but rather it works successfully and numerical analysis of the exact solutions of field equations coincide with all most important observational data particularly late-time-accelerated expansion. So, this new term may be added to the gravitational actions of  $f(T)$-gravity.\\
\end{abstract}

\maketitle
\section{\bf{Introduction} \label{Intro}}
Various astronomical and cosmological observations of the last decade, including CMB studies \cite{1}, supernovae \cite{2,3} and large scale structure \cite{4}, have provided a picture of the universe with accelerating expansion. This profound mystery leads us to the prospect that, either about $70\%$ of the universe is made up of a substance known as dark energy\cite{5}, about which we have almost no knowledge at all, or that General Relativity (GR) is modified at cosmological scales \cite{6,7,8}. A simple candidate for the dark energy is the cosmological constant with the equation of state (EoS) parameter $\omega = -1$. However, the cosmological constant model is subject to the fine-tuning and coincidence problems \cite{9}. In order to solve these problems, various dynamical dark energy models have been proposed consisting quintessence \cite{10,11}, phantom \cite{12,13} and quintom \cite{14,15,16}. Since the quintessence type of matter could not give the possibility that $\omega < -1$, so the extended paradigms (\textit{i.e.} phantom and quintom) are proposed \cite{17}. Beside this unknown-nature dark energy, a second way which is about various gravitational modification theories like $f(R)$, $f(T)$ and scalar-tensor have been lionized. One of the modifications of the matter part of the Einstein-Hilbert action is $f(T)$ gravity as an extension of teleparallel gravity. Teleparallel Gravity (TG), demonstrably equivalent to general relativity, was initially introduced by Einstein for the sake of unifying the gravity and the electromagnetism. In TG we use Weitzenböck connection instead of the Levi-Civita connection, so we have torsion in lieu of curvature only. The field equations in this theory are second-order differential equations, while for the generalized $f(R)$ theory they are fourth-order, thus it is simpler to analyze and elaborate the cosmic evolution \cite{18}. \\

The actions of this context are likely to contain several scalar fields, but it is normally assumed that only one of these fields remains dynamical for a long period, eye-catchingly. We always see the coupling function of $f(R)$ and also $f(T)$ in the form of a function; that is, $F(\varphi) f(R)$ or $F(\varphi) f(T)$, depending on the scalar field only. The motivation for the non-minimal coupling, $F(\varphi)R$ in which $F(\varphi)= \frac{1}{2} \left(\frac{1}{8\pi G} - \xi \varphi^2 \right)$, in the gravitational Lagrangian comes from many directions. However, this explicit non-minimal coupling was originally introduced in the context of classical radiation problems \cite{19}, and also it is requested by renormalizability in curved space-time \cite{20}. For different amounts of $\xi$, we have the following table.
\begin{table*}
\begin{center} \textbf{Table - 1}\end{center}
\begin{center}
\begin{tabular}{|c|c|c|c|c|}
  \hline
  Amounts of $\xi$ & $\xi = 1/6$ & $\xi = 0$ & $\| \xi \| \gg 1$ & (general case) $\xi \neq 0$ \\\hline
  $\quad$ Named as ... coupling $\quad$ & $\quad$ conformal $\quad$ & $\quad$ minimal $\quad$ & $\quad$ strong $\quad$ & $\quad$ (standard) non-minimal $\quad$ \\
  \hline
\end{tabular}\\
\end{center}
\end{table*}
\noindent\\
However, the values of $\xi$ in renormalizable theories depend on the class of theory \cite{56,57}. A nonzero $\xi$ is generated by first loop corrections even if it is absent in the classical action \cite{21,22}. A non-minimal coupling term is expected at high curvatures \cite{23}, and it has been argued that classicalization of the universe in quantum cosmology indeed requires $\xi \neq 0$. Moreover, the non-minimal coupling can solve potential problems of primordial nucleosynthesis \cite{24} and the absence of pathologies in the propagation of $\varphi$-waves seems to require conformal coupling for all non-gravitational scalar fields \cite{25}. Any attempt to formulate quantum field theory on a curved space-time necessarily leads to modifying the Hilbert-Einstein action. This means adding terms containing non-linear invariants of the curvature tensor or non-minimal couplings between matter and the curvature originating in the perturbative expansion \cite{26,27}.\\

Now, let us take the following incomplete action
\begin{equation*}
S = \int d^4x \sqrt{-g} \left[F(\varphi)R+... \right].
\end{equation*}
into account. Eliminating the accelerating term by integral by parts, the corresponding point-like Lagrangian reads
\begin{equation}\label{ll}
L = 6 a \dot{a}^2 F - 6K a F+\textcolor[rgb]{0.33,0.33,1.00}{\underline{\underline{\textcolor[rgb]{0.00,0.00,0.00}{6 a^2 \dot{a} F^{\prime} \dot{\varphi}}}}}+...,
\end{equation}
where $K=0, \pm1$. Here we assume the signature $(+$$-$$-$$-)$ for the FRW's metric components. On the other hand, in $f(T)$-gravity, pursuant to torsion's form, we have no accelerating term, so in this case, we have no the last term in (\ref{ll}) in which derivative of the scalar field couples with scale factor and its derivative. Maybe, it is worth to note what happens when we insert the term as $U\left(\varphi , \nabla_{\mu}\varphi \nabla^{\mu}\varphi \right) g(T)$ in the actions of \textit{f(T)}-gravity context. As mentioned in the first paragraph of the introduction, the ``Teleparallel'' equivalent of General Relativity (TEGR). Altogether, in many cases, the authors construct the actions of $f(T)$-gravity via replacing the torsion insisted of curvature (for example, see \cite{41,a1}). However, when the non-minimal coupling is switched on the resulting theory exhibits different behavior. Hence, the last term in (\ref{ll}) is the inspiring for adding such term. The main purpose of the present work is to answer the aforementioned question by having recourse to the Noether symmetry approach.\\

Symmetries play a substantial role in the theoretical physics. It can safely be said that Noether symmetries are a powerful implement both to select models at a fundamental level and to find exact solutions for specific Lagrangians. In the literature, applications of the Noether symmetry in generalized theories of gravity have been superabundantly studied (for example see \cite{28,29,30,31,32,33,34,35,36,37,38,39,40,41,42,43,44,45,46,47,48,49,50}). Beside this useful approach, another lucrative approach as B.N.S. approach has recently been innovated \cite{51}. B.N.S. approach may carry more conserved currents than Noether symmetry approach. Furthermore, sometimes Noether symmetry approach lacks achieving the purpose. In such cases, utilizing the B.N.S. approach is hobson's choice. Also, with this new procedure, solving process of ordinary differential equation's system, comprised of field equations and conserved currents, is a paved road.\\

The Noether theorem states that for a given lagrangian $L$, defined on the tangent space of configurations $TQ\equiv \{q_{i}, \dot{q}_{i} \}$, if the Lie derivative of the Lagrangian $L$, dragged along a vector field $\textbf{X}$,
\be \label{X}
\textbf{X} = \alpha^{i}(q) \frac{\partial}{\partial q^{i}} +\dot{\alpha}^{i}(q) \frac{\partial}{\partial \dot{q}^{i}},
\ee
where dot means derivative with respect to $t$, vanishes \cite{52}
\be \label{LX}
\mathfrak{L}_{\textbf{X}}L = \textbf{X}_{\mu}L^{\mu} = \alpha^{i}(q) \frac{\partial L}{\partial q^{i}} +\dot{\alpha}^{i}(q) \frac{\partial L}{\partial \dot{q}^{i}}= 0,
\ee
then $\textbf{X}$ is a symmetry for the dynamics and it generates the following conserved quantity (constant of motion)
\be \label{I}
\Sigma_{0} = \alpha^{i} \frac{\partial L}{\partial \dot{q}^{i}}.
\ee
Alternatively, utilizing the Cartan one–form
\be \label{teta}
\theta_{L} \equiv \frac{\partial L}{\partial \dot{q}^{i}} dq^{i}
\ee
and defining the inner derivative
\be \label{inner}
i_{\textbf{X}} \theta_{L} = \langle \theta_{L} , \textbf{X}\rangle
\ee
we get,
\be \label{cartan}
i_{\textbf{X}} \theta_{L} = \Sigma_{0},
\ee
provided that (\ref{LX}) holds. The Eq. (\ref{cartan}) is coordinate independent. Using a point transformation, the vector field $\textbf{X}$ is rewritten as
\be \label{also lift}
\tilde{\textbf{X}} = \left(i_{\textbf{X}} dQ^{k} \right) \frac{\partial}{\partial Q^{k}} + \left[ \frac{d}{dQ^{k}} \left(i_{\textbf{X}} dQ^{k} \right) \right] \frac{\partial}{\partial \dot{Q}^{k}}.
\ee
If $\textbf{X}$ is a symmetry, so is $\tilde{\textbf{X}}$ (\textit{i.e.} $\tilde{\textbf{X}} L =0$), and a point transformation is chosen such that
\be \label{point equ}
i_{\textbf{X}} dQ^{1} = 1 , \qquad i_{\textbf{X}} dQ^{i} = 0 \qquad (i \neq 1).
\ee
It follows that
\be \label{tilde}
\tilde{\textbf{X}} = \frac{\partial}{\partial Q^{1}} , \qquad \frac{\partial L}{\partial Q^{1}} = 0,
\ee
therefore, $Q^{1}$ is a cyclic coordinate and the dynamics can be reduced. However, the change of coordinates is not unique and a clever choice would be advantageous \cite{53}.\\

The structure of the paper is as follows. In section (\ref{II}) we introduce the model and extract the point-like lagrangian and field equations. In section (\ref{III}) we present the Noether symmetries, invariants and exact solutions of the model. Moreover, by data analysis, we demonstrate that the observational data corroborate our findings. In section (\ref{V}) we sum up the obtained graceful results.

\section{The model \label{II}}
Regarding the mentioned points in the second and third paragraphs of the introduction (\ref{Intro}), we want to investigate the following gravitational action in extended gravity context
\be \label{action}\begin{split}
S = \int d^4x e \bigg[ f(\varphi)T& \textcolor[rgb]{0.60,0.20,0.80}{\underbrace{\textcolor[rgb]{0.00,0.00,0.00}{\text{$-$ } U\left(\varphi ,\varphi_{,\mu}\varphi^{,\mu} \right) g(T)}}_{\textbf{The Unusual Term}}}\\ \\& - \frac{\omega(\varphi)}{2}\varphi_{,\mu} \varphi^{,\mu} + V(\varphi) \bigg],
\end{split}\ee
where $e=\det(e_{\nu}^{i})=\sqrt{-g}$ with $e_{\nu}^{i}$ being a vierbein (tetrad) basis, $f(\varphi)$ is the generic function describing the coupling between the scalar field and scalar torsion $T$, $\varphi_{,\mu}$ indicates the covariant derivative of $\varphi$, $U\left(\varphi ,\varphi_{,\mu}\varphi^{,\mu} \right)$ is the unknown coupling function which we hypothesize it, in general, to depend on scalar field and gradients of it. This function coupled with an unknown function of torsion $g(T)$. Here, $\omega(\varphi)$ and $V(\varphi)$ are the coupling function and scalar potential, respectively. Note that the scalars here are caused by conformal symmetry \cite{58}. We presume that the geometry of space–time is described by the flat FRW metric which is consistent with the present cosmological observations
\begin{equation}\label{metric}
ds^2 = dt^2 - a^2(t) \left[dx^2 + dy^2 + dz^2\right],
\end{equation}
where the scale factor $a$ is a function of time. With this background geometry, the scalar torsion takes the form $T=-6\dot{a}^2/{a^2}$. First, for simplifying the action, we set the following form by assuming that two main parts of $U$ are separable
\begin{equation}\label{forms}
U\left(\varphi ,\varphi_{,\mu}\varphi^{,\mu} \right) =h(\varphi) \Phi (\dot{\varphi})  = h(\varphi) \dot{\varphi},
\end{equation}
where $h(\varphi)$ is an unknown function of the scalar field $\varphi$, and the dot represents a differentiation with respect to $t$. We can not present any physical argument behind such choice for the unknown function $\Phi (\dot{\varphi})$, rather relies on the fact that it works fairly and the Hessian determinant turns out to be zero through this choice of function after finding the other unknown coupling functions via Noether approach. Moreover, speaking of the last term in (\ref{ll}) and mentioned points in the introduction, it is better we fit the second main part as $\dot{\varphi}$ at first. Using (\ref{action}), (\ref{forms}) and the Lagrange's method of undetermined coefficients, the action (\ref{action}) can be written as
\begin{equation}\label{action2}\begin{split}
S = \int d^4x e \bigg[ &f(\varphi)T - h(\varphi)\dot{\varphi} g(T)\\& -\lambda \left(T + 6 \frac{\dot{a}^2}{a^2} \right) - \frac{\omega(\varphi)}{2} \dot{\varphi}^2 + V(\varphi) \bigg],
\end{split}\end{equation}
where the Lagrange multiplier $\lambda$ is derived by varying the action (\ref{action2}) with respect to $T$
\begin{equation}\label{lambda}
\lambda = f - h\dot{\varphi}g^{\tau},
\end{equation}
in which the $\tau$ denotes a differentiation with respect to the torsion $T$. So, the point-like Lagrangian corresponding to the action (\ref{action}) becomes
\begin{equation}\label{point like Lagrangian}
L = fTa^{3}-h\dot{\varphi}ga^3- (f-h\dot{\varphi}g^{\tau})[Ta^3+6\dot{a}^2a]-\frac{1}{2}\omega \dot{\varphi}^2 a^3+Va^3.
\end{equation}
Hence, the Euler-Lagrange equations for the scale factor $a$ would be
\begin{equation}\label{Fe 1}\begin{split}
&4\left(\frac{\ddot{a}}{a}\right) \left(f - h \dot{\varphi} g^{\tau} \right) + 2\left(\frac{\dot{a}^2}{a^2} \right)^2 \left(f - h\dot{\varphi} g^{\tau} \right) \\&+ 4 \left(\frac{\dot{a}}{a} \right) \left(f^{\prime}\dot{\varphi}-hg^{\tau} \ddot{\varphi}-hg^{\tau \tau} \dot{\varphi}\dot{T}-h^{\prime}g^{\tau}
\dot{\varphi}^2 \right)\\&+\left(Tg^{\tau}-g \right)h\dot{\varphi}-\frac{1}{2}\omega \dot{\varphi}^2 +V=0,
\end{split}\end{equation}
where the prime indicates a derivative with respect to $\varphi$. For the scalar field, $\varphi$, the Euler-Lagrange equation takes the following form
\begin{equation}\label{Fe 2}\begin{split}
&12h g^{\tau} \left(\frac{\ddot{a}}{a} \right)\left(\frac{\dot{a}}{a} \right)+6h g^{\tau} \left(\frac{\dot{a}}{a} \right)^3 +6 \left(\frac{\dot{a}}{a} \right)^2 \left(f^{\prime} +h g^{\tau \tau} \dot{T} \right)\\&-3\left(\frac{\dot{a}}{a} \right) \left(hg-h g^{\tau}T+\omega \dot{\varphi} \right) +h g^{\tau \tau}T \dot{T} - \omega \ddot{\varphi}-\frac{1}{2}\omega^{\prime} \dot{\varphi}^2 - V^{\prime} =0,
\end{split}\end{equation}
which is the Klein-Gordon equation. The energy function which is the $\binom{0}{0}$-Einstein equation, associated with the point-like Lagrangian (\ref{point like Lagrangian}) is found as
\begin{equation}\label{Fe 3}
\left(12h g^{\tau} \dot{\varphi} -6f \right) \left( \frac{\dot{a}}{a}\right)^2 -\frac{1}{2}\omega \dot{\varphi}^2 -V=0.
\end{equation}
And, the Euler-Lagrange equation for the torsion scalar $T$ reads
\begin{equation}\label{Fe 4}
a^3 h \dot{\varphi}g^{\tau \tau} \left(T + \frac{\dot{a}^2}{a^2} \right) =0.
\end{equation}
From Eq. (\ref{Fe 4}), there are three possibilities: (1) $g^{\tau \tau} = 0$ implying linearity for $g(T)$ which is not interesting, for we have linear form of torsion in the action (\ref{action}), (2) $\dot{\varphi}=0$ which leads to the linear form for the scalar field, so it is not suitable and (3) the possibility of
\begin{equation*}
T = - 6 \frac{\dot{a}^2}{a^2},
\end{equation*}
which is the definition of scalar torsion for flat FRW.
\section{Exact Solutions Via Noether Symmetry Approach and Data Analysis\label{III}}
In this section, we utilize the Noether symmetry approach for solving Eqs. (\ref{Fe 1})-(\ref{Fe 4}). The configuration space of the point-like Lagrangian (\ref{point like Lagrangian}) is $Q=\{a, \varphi , T\}$ whose tangent space is $TQ = \{a, \dot{a}, \varphi, \dot{\varphi}, T, \dot{T}\}$. The existence of the Noether symmetry implies the existence of a vector field $\textbf{X}$,
\begin{equation}\label{X}
\textbf{X} = \alpha \frac{\partial}{\partial a}+\beta \frac{\partial}{\partial \varphi} +\gamma \frac{\partial}{\partial T} + \alpha_{,t} \frac{\partial}{\partial \dot{a}}+\beta_{,t} \frac{\partial}{\partial \dot{\varphi}}+\gamma_{,t} \frac{\partial}{\partial \dot{T}},
\end{equation}
where
\begin{equation*}\begin{split}
y =& y (a, \varphi, T)\\&\longrightarrow \quad y_{,t} = \dot{a} \frac{\partial y}{\partial a}+ \dot{\varphi} \frac{\partial y}{\partial \varphi}+\dot{T} \frac{\partial y}{\partial T} \quad ; \quad y \in \{\alpha, \beta, \gamma\},
\end{split}\end{equation*}
such that
\begin{equation*}\begin{split}
&\mathfrak{L}_{\textbf{X}}L =0\\& \longrightarrow  \alpha \frac{\partial L}{\partial a}+\beta \frac{\partial L}{\partial \varphi} + \gamma \frac{\partial L}{\partial T}+ \alpha_{,t} \frac{\partial L}{\partial \dot{a}}+\beta_{,t} \frac{\partial L}{\partial \dot{\varphi}} +\gamma_{,t} \frac{\partial L}{\partial \dot{T}} =0.
\end{split}\end{equation*}

This condition yields the following system of linear partial differential equations
\begin{equation}\label{NS-Eq-1}
\frac{\partial \alpha}{\partial b} = 0, \quad \frac{\partial \alpha}{\partial T} = 0, \quad \frac{\partial \beta}{\partial a} =0, \quad \frac{\partial \beta}{\partial T} = 0,\\
\end{equation}
\begin{equation}\label{NS-Eq-2}\begin{split}
&3 \alpha V+\beta a V^{\prime}=0, \qquad 2af \left(\frac{\partial \alpha}{\partial a}\right)+\alpha f+a\beta f^{\prime} = 0,\\&2a \omega \left(\frac{\partial \beta}{\partial \varphi}\right)+3 \alpha \omega +a \beta \omega^{\prime} = 0,\\
\end{split}\end{equation}
\begin{equation}\label{NS-Eq-3}\begin{split}
6 \alpha h g^{\tau}+ 6a\beta h^{\prime} g^{\tau}+&12ah g^{\tau} \left(\frac{\partial \alpha}{\partial a}\right)\\&+6hag^{\tau} \left(\frac{\partial \beta}{\partial \varphi}\right)+6ah \gamma g^{\tau \tau}=0,\\
\end{split}\end{equation}
\begin{equation}\label{NS-Eq-4}\begin{split}
\left(aTh g^{\tau}-ahg \right) \left(\frac{\partial \beta}{\partial \varphi}\right)&+aTh \gamma g^{\tau \tau}+3\alpha Th g^{\tau}\\&+\beta aT g^{\tau} h^{\prime}-3\alpha hg-\beta h^{\prime}ga =0.
\end{split}\end{equation}
This system of linear partial differential equations can be solved by using the separation of variables. Hence, one may obtain
\begin{equation}\begin{split} \label{sol for NS}
&f(\varphi)=f_{0}\varphi^n, \quad h(\varphi)=h_{0}\varphi^{n-1}, \quad V(\varphi)= V_{0} \varphi^{n},\\ &g(T)=(-6 T)^{1/n},\quad
\omega (\varphi)= \omega_{0} \varphi^{n-2},  \quad \alpha =\alpha_{0} a,\\& \beta= \beta_{0} \varphi, \quad \gamma = 0,
\end{split}\end{equation}
in which $f_{0}$, $\omega_{0}$, $V_{0}$, $h_{0}$, $\alpha_{0}$, and $\beta_{0}$ are constants of integration and $\beta_{0}=-3\alpha_{0}/n$ and $h_{0}=nf_{0}$. This solution holds for $n = 2$, only. Of course, $n=2$ has a physical nature as well, because the common form of $f(\varphi)$ can be given by the non-minimal coupling as $f(\varphi) = \frac{M^2}{2} \left(\frac{1}{\kappa} - \xi \varphi^2 \right)$, however, certain Grand-Unified theories lead to a polynomial coupling of the form $1+\xi \varphi^2+\zeta \varphi^4$, but on the other hand $\omega$ must be dimensionless, so $n$ should be equal to $2$. According to (\ref{sol for NS}), symmetry generator turns out to be
\begin{equation}\label{sym. gen.}
\textbf{X} = a \frac{\partial}{\partial a}+ \frac{-3}{2} \varphi \frac{\partial}{\partial \varphi}+\dot{a}\frac{\partial}{\partial \dot{a}}+\frac{-3}{2} \dot{\varphi}\frac{\partial}{\partial \dot{\varphi}}.
\end{equation}
Hence corresponding conserved current is found as
\begin{equation}\label{Noether current}
\textbf{I} = 3 \alpha_{0} a^2 \varphi \left(2f_{0}\varphi \dot{a}-4 f_{0}a\dot{\varphi}+\frac{\omega_{0}}{2}a\dot{\varphi} \right).
\end{equation}
Since the form of $g(T)$ has been specified (\ref{sol for NS}) and $\gamma =0$, and on the other hand, regarding the third option in Eq. (\ref{Fe 4}), it is an ineffective shot to tow $T$ in $Q$. Therefore, the configuration space reduces to two $Q=\{a, \varphi\}$. This means we can rewrite the point-like Lagrangian (\ref{point like Lagrangian}) free of $T$ by substituting the form of torsion $T$. Now, by assuming that $\textbf{X}$ is a symmetry, we seek a point transformations on the vector field $\textbf{X}$ such that
\begin{equation}\label{point transformation}
i_{\textbf{X}}dz =1, \qquad \qquad i_{\textbf{X}} dp=0,
\end{equation}
whereas $i: (a,\varphi) \rightarrow (z,p)$ in which $z=z(a,\varphi)$ and $p=p(a,\varphi)$. Here $z$ is a cyclic variable. If we were to keep $T$, then it had to be mapped to itself, i.e. $i: (a,\varphi, T) \rightarrow (z, p, T)$. Solving the Eq. (\ref{point transformation}) leads to
\begin{equation}\label{sol for cyc.}
p=a^{3/2} \varphi, \qquad \qquad z=a^{3/2} \varphi + \ln(a).
\end{equation}
So, the corresponding inverse transformation is
\begin{equation}\label{inv. trans.}
a(p,z)=\exp(z-p), \qquad \varphi(p,z)=p \exp\left(\frac{3}{2} (p-z) \right).
\end{equation}
It is clear that (\ref{sol for cyc.}) is an arbitrary choice since more general conditions are possible. The point-like Lagrangian (\ref{point like Lagrangian}) can be rewritten in terms of cyclic variables by (\ref{inv. trans.}) as
\begin{equation}\label{re-point like}
\mathcal{L }=\frac{\dot{p}^2}{2} \left(\frac{3}{4}p+1 \right)-p \left(\frac{3}{8}\dot{p} \dot{z}-V_{0}p \right),
\end{equation}
in which we put $\omega_{0}=1$ and $f_{0} = 3/32$ or equivalently $h_{0}=3/16$. The Euler-Lagrange equations relevant to (\ref{re-point like}) are
\begin{equation}\begin{split}\label{re-Fe 1}
\frac{d}{dt} \left(p \dot{p} \right) =0 \qquad; \qquad \frac{d}{dt} \left(p \dot{p} \right)\equiv \textbf{I}
\end{split}\end{equation}
\begin{equation}\begin{split}\label{re-Fe 2}
\frac{3}{8} p \ddot{z}-\ddot{p} \left(1+\frac{3}{4}p \right)-\frac{3}{8}\dot{p}^2 -2 V_{0}p=0.
\end{split}\end{equation}
Note that Eq. (\ref{re-Fe 1}) is equivalent to $\textbf{I}$ which is given by Eq. (\ref{Noether current}), so rewriting Eq. (\ref{Noether current}) in terms of cyclic variables does not add a new equation. And Hamiltonian constraint reads
\begin{equation}\label{re-hamil.}
\frac{1}{2}\dot{p}^2 \left(\frac{3}{4}p+1 \right)-\frac{3}{8}\dot{z} \dot{p} p + V_{0} p^2=0.
\end{equation}
One can write $const.$ in right side of Eq. (\ref{re-hamil.}) instead of zero, as in general, we have $E_{L}=const.$. Solving Eqs. (\ref{re-Fe 1})-(\ref{re-Fe 2}) through (\ref{re-hamil.}) leads to
\begin{equation}\label{sol. cys. p}
p(t)= \sqrt{2\left(c_{1}t+c_{2} \right)},\\
\end{equation}
\begin{equation}\label{sol. cys. z}\begin{split}
z(t) =  \frac{2}{3} &\ln\left(\left|c_{1}t+c_{2} \right| \right)+\frac{8}{3}V_{0}t^2 \\& +\left(c_{3}-\frac{16}{3}\frac{V_{0}c_{2}}{c_{1}} \right)t+\sqrt{2\left(c_{1}t+c_{2} \right)} +c_{4},
\end{split}\end{equation}
where $\{c_{i}$ ; $i = 1,...,4\}$ are constants of integration. Doing inverse transformations by the use of Eq. (\ref{inv. trans.}) give the solutions of Eqs. (\ref{Fe 1})-(\ref{Fe 3}) and $\textbf{I}=0$ (See Eq. (\ref{Noether current})) as
\begin{figure*}
\centering
\includegraphics[width=7 in, height=2.7 in]{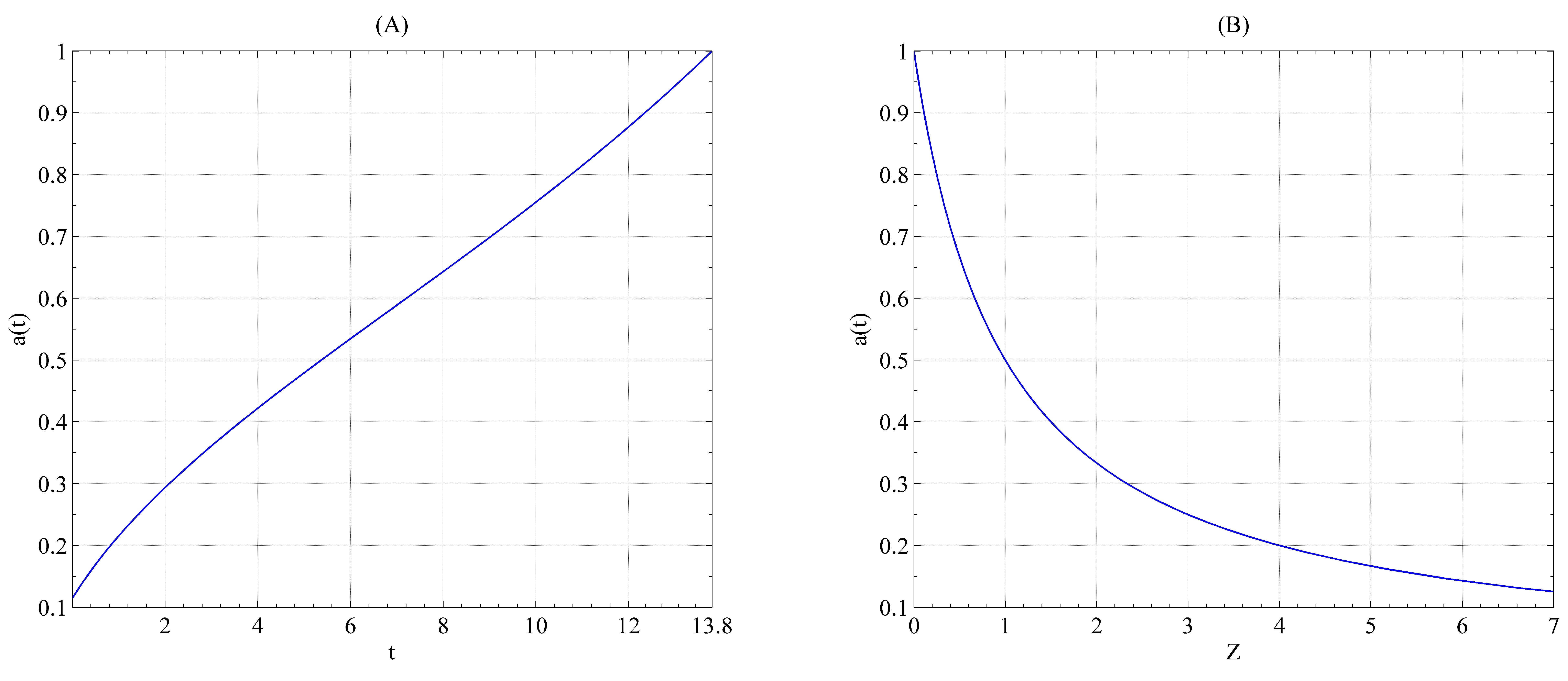}\\
\caption{Plot (A) indicates the scale factor $a(t)$ versus time $t$ at the time range $ [0.0006 , 13.816]$ while plot (B) shows the scale factor $a(t)$ versus redshift $z$ at the redshift range $[0 , 7]$.}\label{fig1}
\end{figure*}
\begin{figure*}
\centering
\includegraphics[width=7 in, height=2.7 in]{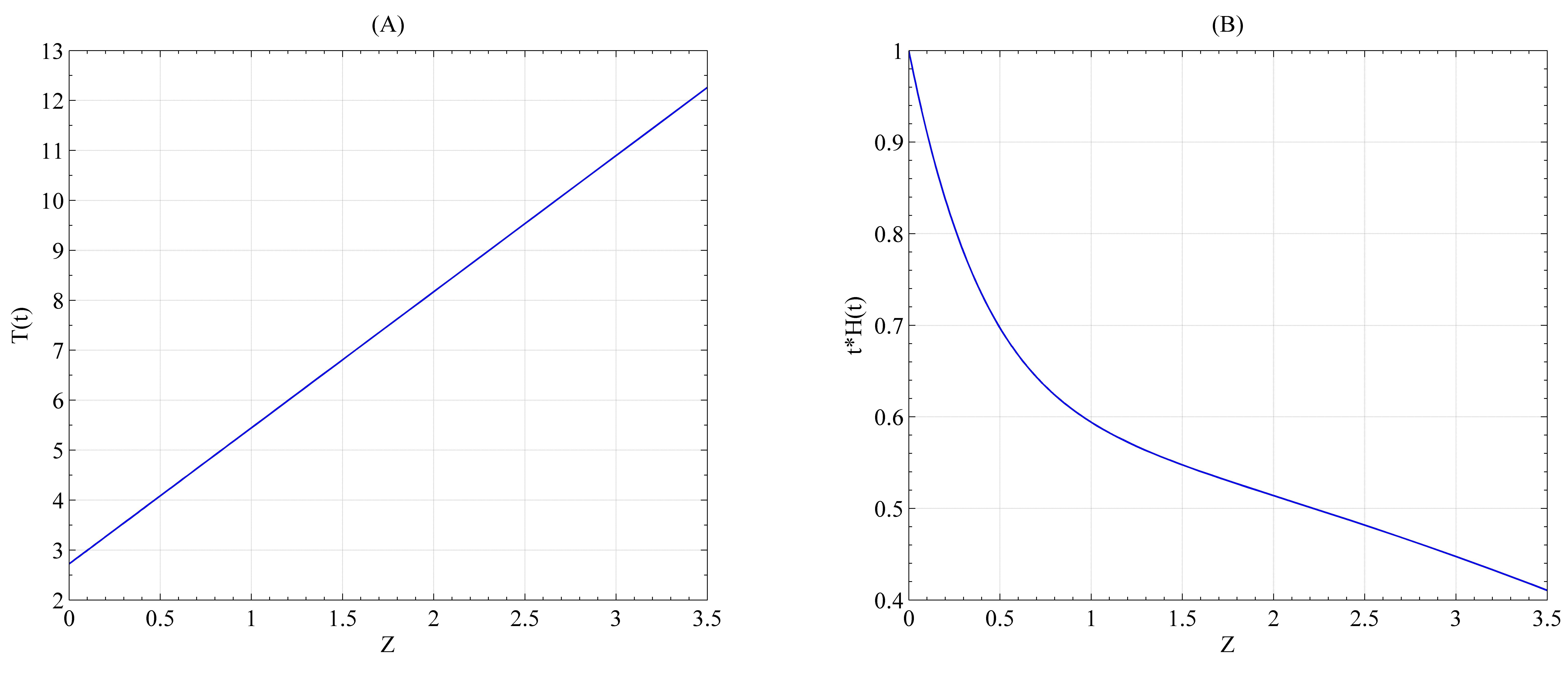}\\
\caption{Plots (A) and (B) indicate the temperature $T(t)$ and  $ t \times H(t)$ versus redshift $z$ at the redshift range $[0 , 3.5]$, respectively.}\label{fig2}
\end{figure*}
\begin{figure*}
\centering
\includegraphics[width=7 in, height=2.7 in]{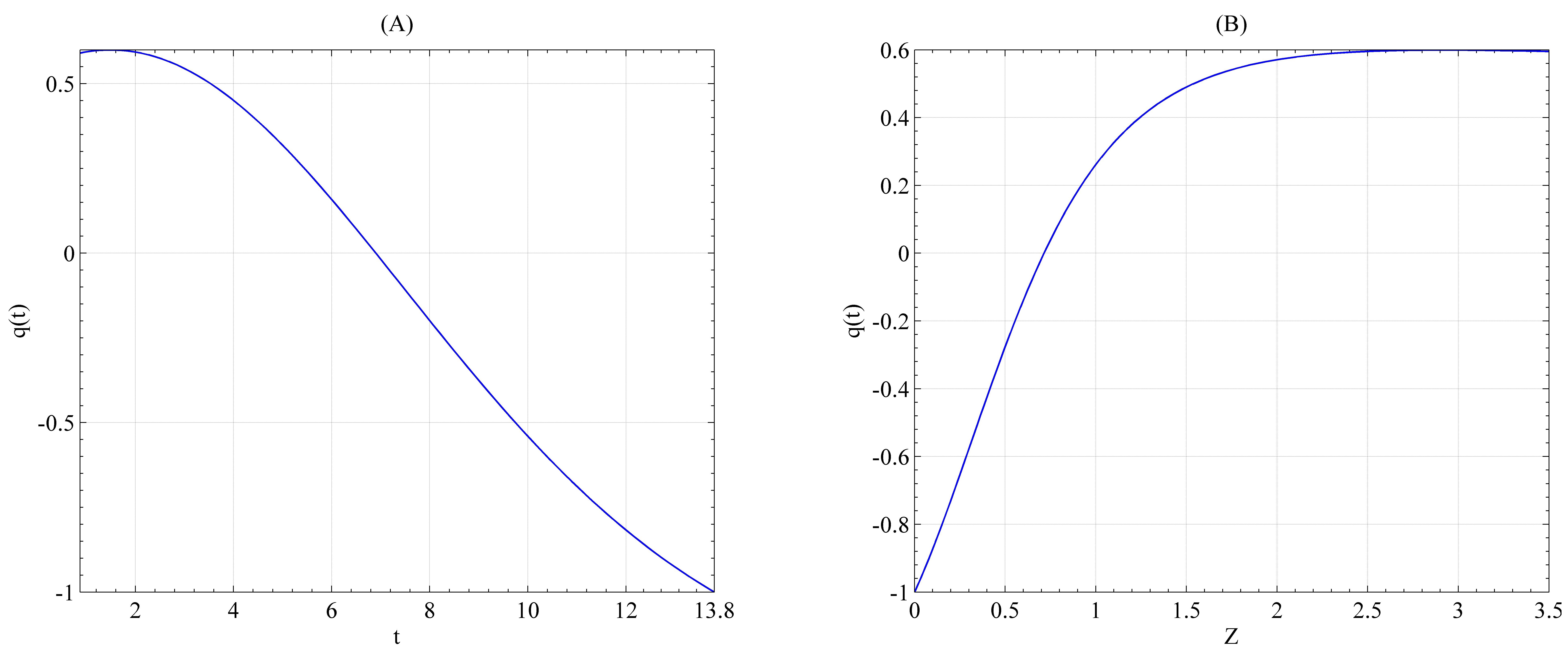}\\
\caption{Plot (A) indicates the deceleration parameter $q(t)$ versus time $t$ at the time range $[1 , 13.816]$ while plot (B) shows the deceleration parameter $q(t)$ versus redshift $z$ at the redshift range $[0,3.5]$.}\label{fig3}
\end{figure*}
\begin{figure*}
\centering
\includegraphics[width=7 in, height=2.7 in]{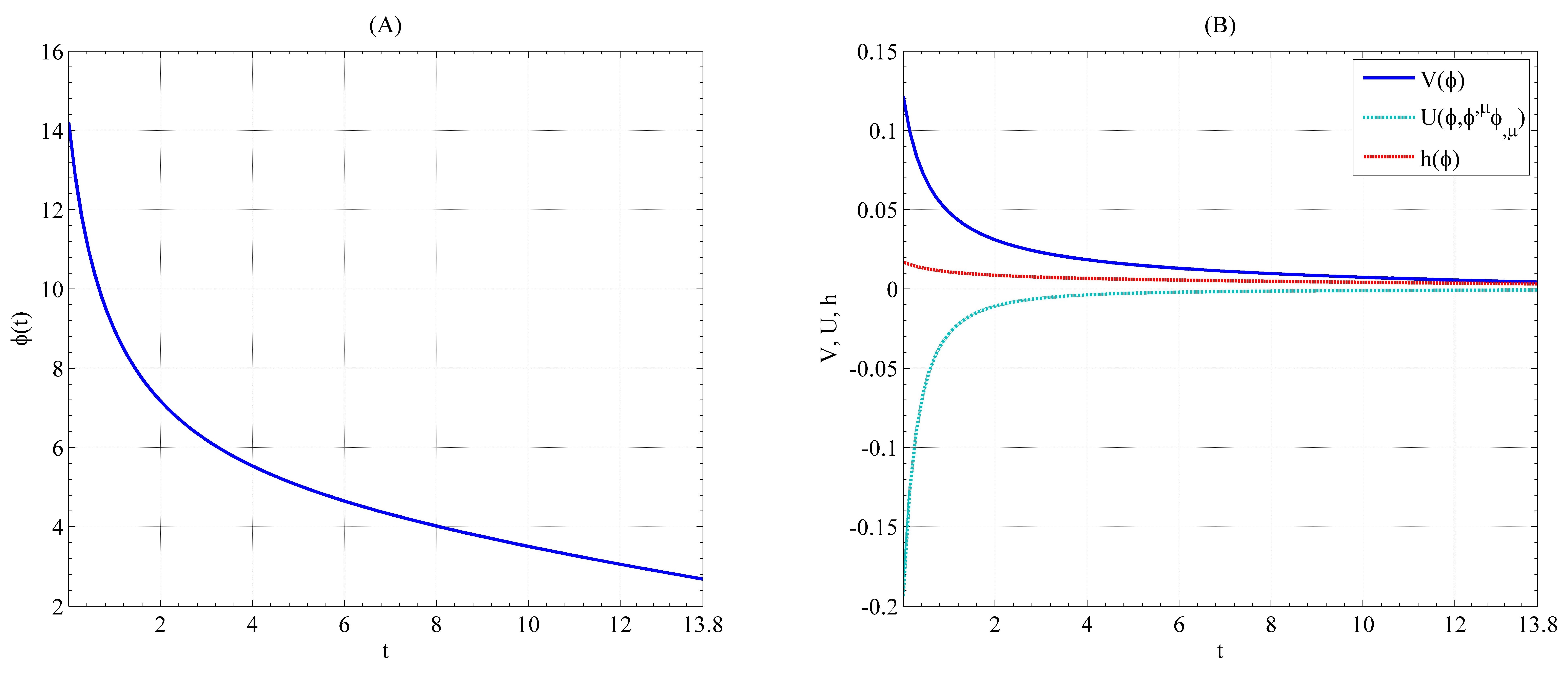}\\
\caption{Plot (A) indicates the scalar filed $\varphi(t)$ versus time $t$ at the time range $[0.0006 , 13.816]$ and plot (B) shows coupling function $U(\varphi, \varphi_{\mu} \varphi^{\mu})$, $h(\varphi)$ and the scalar potential $V(\varphi)$ versus time $t$ at the same time range.}\label{fig4}
\end{figure*}
\begin{figure*}
\centering
\includegraphics[width=3.5 in, height=2.7 in]{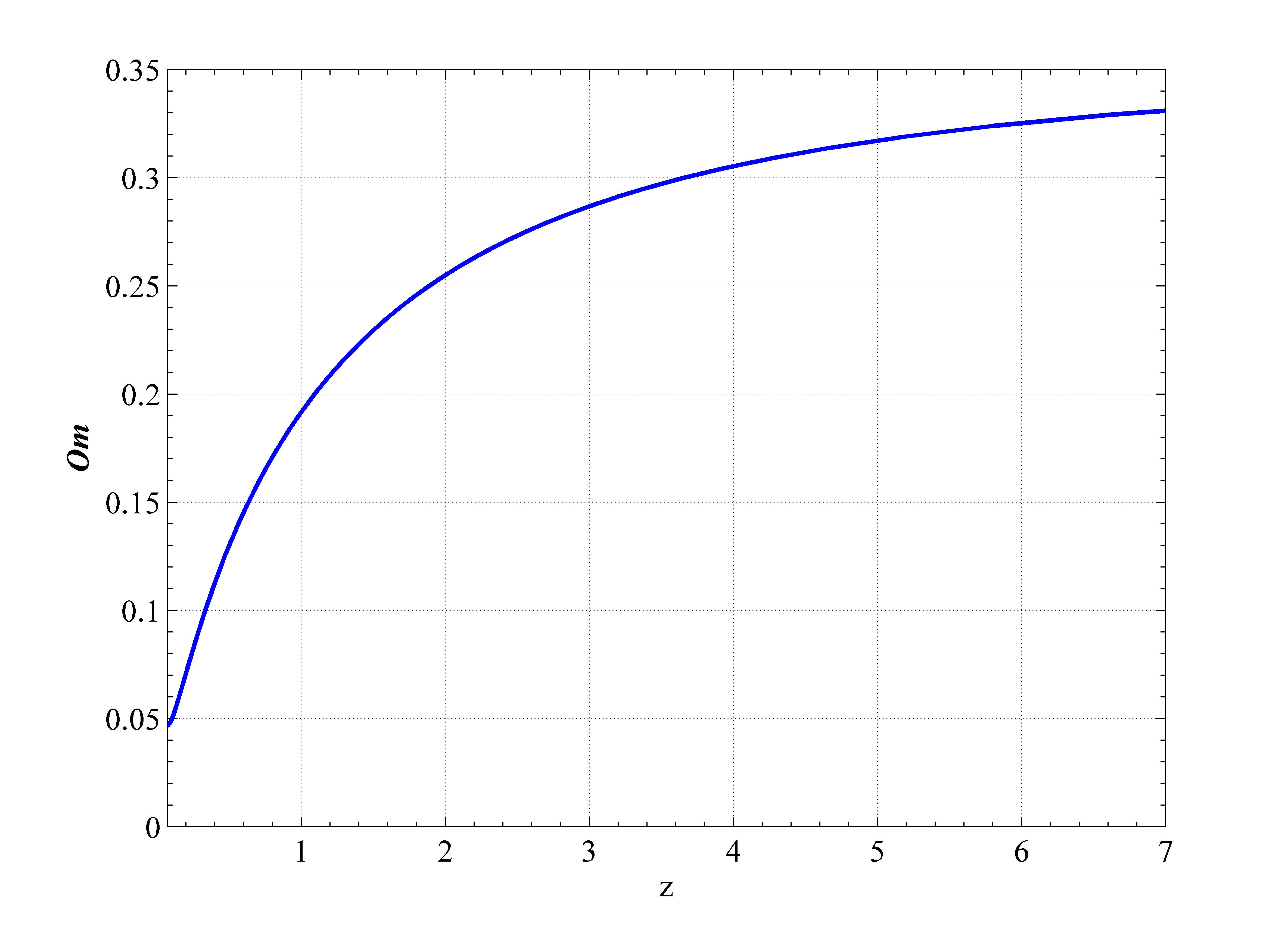}\\
\caption{Plot indicates the \textit{Om}-diagnostic versus redshift $z$ at the redshift range [0.07 , 7].}\label{fig5}
\end{figure*}

\begin{equation}\label{solutions-a}
a(t)=\left(c_{1} t+c_{2}\right)^{2/3} \exp\left(\frac{8}{3}V_{0}t^2 +\left[c_{3}-\frac{16}{3}\frac{V_{0}c_{2}}{c_{1}} \right]t+c_{4}\right),\\
\end{equation}
\begin{equation}\label{solutions-varphi}
\varphi(t) = \frac{\exp\left(-4V_{0} t^2 - \frac{3}{2} \left[c_{3}-\frac{16}{3} \frac{V_{0}c_{2}}{c_{1}}\right]t-\frac{3}{2}c_{4} \right)}{\sqrt{2(c_{1}t+c_{2})}}.
\end{equation}
Therefore, these solutions carry two conserved currents; $E_{L}$ (Eq. (\ref{Fe 3})) and $\textbf{I}$ (Eq. (\ref{Noether current})) that correspond to two symmetry generators $\textbf{X}_{1}=\partial / \partial t$ and $\textbf{X}$ (Eq. (\ref{sym. gen.})), respectively.\\

For illustrating the descriptions of late-time-accelerated expansion from the perspective of the studied model, we single out the constants as below
\begin{equation}\label{selections}\begin{split}
&c_{1} = 1, \quad c_{2} = 0.6068, \quad  c_{3} = -0.0162, \\& c_{4} = - 1.8340, \quad V_{0} = 0.0006.
\end{split}\end{equation}
We present five figures with data analysis with time unit 1 Gyr $\equiv$ $\mathbf{1}$. The behavior of the scale factor $a$ versus time and redshift in figure \ref{fig1} imply; first, the scale factor with increasing nature expressing initially the decelerated and then accelerated expansion of the universe (See figure \ref{fig1}(A)), second, the scale factor versus redshift plot \ref{fig1}(B), confirms that the present value of the scale factor is exactly $1$, so, from figure \ref{fig1}(A) we learn that the age of universe is $t_{0} = 13.816$ Gyr. As usual, ignoring small variation of the prefactor, we consider that the CMBR temperature falls as $a^{-1}$, then, the present value of it is, $T_{0} = 2.725$ K (See figure \ref{fig2}(A)). The evolution of the Hubble parameter (not presented) gives its present value, $H_{0} = 7.238 \times 10^{-11}$ yr$^{-1}$ $\equiv 70.82$ Km.s$^{-1}$.Mpc$^{-1}$. It is a nice result, since the recent observational data tell $H_{0} = 72.25 \pm 2.38$ Km.s$^{-1}$.Mpc$^{-1}$ \cite{60}. Therefore $H_{0} \times t_{0} = 1$, which fits the observational data with high precision, as plotted in figure \ref{fig2}(B). Figure \ref{fig3}(A) indicates deceleration parameter, $q = -(a \ddot{a}) / \dot{a}^2$, versus time plot. It shows a pass from negative to positive values which states
first decelerating universe, $q > 0$, then accelerating universe, $q < 0$, and the present value of it, is $q_{0}= -1$, as we expect. Limpidly, $q=0$ renders the inflection point (\textit{i.e.} shifting from decelerated to accelerated expansion in figure \ref{fig1}(A)). So, at redshift $z_{acce.}=0.712$ or equivalently $t_{acce.}=6.198$ Gyr acceleration started (See figure \ref{fig3}(B)). It coincides with the astrophysical data, for observational data concede that at redshift $z_{acce.}<1$, or equivalently at about half the age of the universe acceleration commenced. It is well-known that scalar field must decrease with time. As figure \ref{fig4}(A) shows, our scalar field is consistent with this point. Figure \ref{fig4}(B) indicates the manners of the scalar potential $V(\varphi)$, coupling function $U(\varphi, \varphi_{,\mu}\varphi^{,\mu})$ and $h(\varphi)$ versus time. The behavior of scalar potential is admissible (\textit{i.e.} detractive in face of time). With an eye to action (\ref{action}), we found that the terms $U$ and $V$ have different sign. So, by applying the minus sign of $U$ we learn that both $V$ and $U$ are positive and have subtractive behaviors versus time. A marked difference between those is that $U$ falls sharply than $V$. It is obvious that $h$ has a narrow band around zero, and may this restricts the behavior of $U$. In addition, all these three functions turn out to be almost constants with a little difference from that of the present time. Before terminating this section, we would like to investigate the \textit{Om-diagnostic} analysis. In this manner, we present the figure \ref{fig5} which shows the \textit{Om-diagnostic} parameter versus redshift. The \textit{Om-diagnostic} is an important geometrical diagnostic proposed by Sahni \textit{et al.} \cite{54}, in order to classify the different dark energy (DE) models. The $Om$ is able to distinguish dynamical DE from the cosmological constant in a robust manner both with and without reference to the value of the matter density. It is defined as \cite{55}
\begin{equation}\label{Om}
Om(z) \equiv \frac{\left[\frac{H(z)}{H_{0}}\right]^2 -1}{\left(1+z \right)^3 -1}.
\end{equation}
For dark energy with a constant equation of state (EoS) $\omega$, it reads
\begin{equation}\label{Om-w}
Om(z) = \Omega_{m0} + (1-\Omega_{m0}) \frac{(1+z)^{3(1+\omega)}-1}{(1+z)^{3}-1},
\end{equation}
so, $Om(z) = \Omega_{m0}$ states the $\Lambda$CDM model, therefore the regions $Om(z) > \Omega_{m0}$ and $Om(z) < \Omega_{m0}$ correspond with quintessence ($\omega > -1$) and phantom ($\omega < -1$), respectively. The figure \ref{fig5} indicates phase crossing from quintessence ($Om(z) \gtrsim 0.3$) to phantom ($Om(z) \lesssim 0.3$). Hence, now we are in phantom phase.\\

\section{Conclusion \label{V}}
Owing to the mentioned fact in the introduction (\ref{Intro}), we applied the term $U(\varphi, \varphi_{,\mu}\varphi^{,\mu})g(T)$ within a generic action in $f(T)$-gravity context in FRW background space-time. Whereas the resulting system was overdetermined, so for simplifying the action we did a suitable choice for the unknown function $U=h(\varphi)\dot{\varphi}$ by assuming the separation of two main parts. Then by applying the Noether symmetry approach and using cyclic variables method, we could reach at nice results as our data analysis showed deep compatible results with observational data.\\
In a nutshell, some of our noteworthy findings were,
\begin{enumerate}
  \item The present value of the scale factor is $a_{0} = 1$ (\textit{i.e.} $(z_{0},a_{0})=(0,1)$), so the present value of temperature may be found as $T_{0} = 2.725$ K.
  \item Age of the universe $t_{0} = 13.816$ Gyr.
  \item The resulting model can give Late-time-accelerated expansion. The universe before entering accelerated expansion epoch, became a victim of Friedmann-like matter dominated era for quite a long time. Thus, acceleration dawns at the redshift value $z_{acce.}=0.712$ which is equivalent to $t_{acce.} = 6.198$ Gyr that is about half the age of the universe.
  \item The present value of the Hubble parameter is $H_{0} = 7.238 \times 10^{-11}$ yr$^{-1}$ or equivalently \\$H_{0}=70.82$ Km.s$^{-1}$.Mpc$^{-1}$.\\
  \item $H_{0} \times t_{0} = 1$.
  \item The present value of the deceleration parameter is $q_{0}= -1$.
  \item \textit{Om}-diagnostic shows phase crossing from quintessence to phantom phase.
\end{enumerate}
The added extra function, $U(\varphi, \varphi_{,\mu}\varphi^{,\mu})$, affected like the scalar potential $V(\varphi)$ with the difference that $U$ decayed sharply than $V$. Moreover, around the present time, their amount had negligible difference. \\
Considering the deep compatible results with astrophysical and observational data, we conclude that entering such term, $U\left(\varphi ,\varphi_{,\mu}\varphi^{,\mu} \right) g(T)$, to the scope of the actions of $f(T)$-gravity, not only has no anomalous effect (at least in studied case) but it may give desired results.\\

\section*{Acknowledgment}
I thank Professor Sergei D. Odintsov for useful remarks.\\

\end{document}